\documentclass[aps,pra,twocolumn,showpacs,superscriptaddress,groupaddress]{revtex4-1}

\usepackage{amsfonts,amssymb,amsmath, braket}
\usepackage{mathtools}
\usepackage[]{graphics,graphicx,epsfig}
\usepackage{amsthm,multirow}
\usepackage{color}
\usepackage{diagbox}
\begin{document}

\title{Quantum Radar Cross Section with two-photon entangled states}

\author{Sunghwa Kang}
\email[]{kangsunghwa0514@gmail.com}

\author{Jihwan Kim}

\author{Zaeill Kim}

\author{Duk Y. Kim}

\author{Yong Sup Ihn}

\author{Su-Yong Lee}
\email[]{suyong2@add.re.kr}

\affiliation{Agency for Defense Development, Daejeon 34186, Korea}

\author{\\ Sean Crowe}

\author{Stefan Evans}

\author{Marcio de Andrade}

\author{Joanna Ptasinski}

\affiliation{Cryogenic Electronics \& Quantum Research, Naval Information Warfare Center Pacific, USA}

\date{\today}

\begin{abstract}
We study two-photon entangled states for quantum radar cross section (QRCS), which is an extension of a single-photon QRCS formula. 
Since signal-idler entanglement does not provide any enhancement of the QRCS [Brandsema's PhD Thesis (2017)], we focus on signal-signal entanglement and derive the corresponding biphoton QRCS. We show that it can provide an enhancement over the single-photon QRCS and two-photon separable QRCS, 
where the performance is evaluated for various two-dimensional target geometries in monostatic/bistatic configurations. Furthermore, using the double-Gaussian approximation, we derive QRCS formula for biphoton states with arbitrary degree of entanglement and compute the resulting scattering patterns.
\end{abstract}

\maketitle

\section{Introduction}

Radar cross section (RCS) is the area that we can measure a target by radar, which does not depend on target distance and strength of incident fields but highly on the target geometry and composition \cite{Skolnik}. In a far-field region, it is evaluated with the incident and scattered electric field intensities. 
In quantum regime, by using atom-photon scattering processess, an effective scattering cross-section is quantified as quantum radar cross section (QRCS) \cite{Lanzagorta} whose performance is evaluated with expectation values of the incident and scattered fields. 
QRCS has been studied with single photon states, which shows side-lobe enhancement over the RCS \cite{Brandsema}. The advantge of the QRCS is based on the concept of superposition.
Previously, QRCS was studied with $N$-photon state to detect a rectangular flat plate \cite{Tian:22} as well as with different target geometries and environmental conditions \cite{Hu:23, Tian:24}.
It was also investigated for the polarization effect \cite{brandsema2017effect}, bistatic scenarios \cite{fang2018calculation}, and influence on target detection performance \cite{liu2014analysis}.

QRCS is based on quantum radar \cite{Lanzagorta} that exploits nonclassical states of light to measure remote target properties, such as target range, angles, size, speed, and features. The quantum radar is different from quantum illumination that discriminates the presence or absence of a low-reflectivity target with entanglement in a fixed range \cite{Lloyd,Tan,Kim23}. Several studies have explored on measuring a target range \cite{Maccone,Zhuang,Jeon}. In quantum illumination and target ranging, it is essential to prepare entanglement between signal and idler modes. 
However, the signal-idler entanglement does not provide enhancement of the QRCS \cite{brandsema2017formulation}.

Here, we study entanglement between two signal photons, such that both photons interact with a target in the scattering process. 
We employ the scattering theory of maximally entangled two-photon states \cite{Schotland:16} to derive modified QRCS formula. 
By incorporating two-photon scattering theory into the QRCS formula, we derive a biphoton QRCS formula that differs from single-photon QRCS or separable two-photon QRCS.
Using double-Gaussian approximation \cite{PhysRevA.83.060302, Fedorov_2009}, which assumes a Gaussian pump profile and approximates a joint biphoton amplitude using a double Gaussian function, we derive the joint amplitude as a product of Gaussian envelopes in terms of sum and difference coordinates, where the widths are set by the pump divergence and the phase-matching angular bandwidth of the nonlinear crystal. The ratio of these widths characterizes the degree of entanglement.
With the biphoton QRCS formula, we evaluate the QRCS numerically for various two-dimensional target geometries in monostatic and bistatic configurations.

\section{Two-photon entangled state for QRCS}
QRCS is based on how the incident photon interacts with the atoms in an object.
As an analogy to its classical counterpart, the QRCS is defined as the ratio of the expectation values of the incident and scattered intensities as follows \cite{brandsema2017formulation}:
\begin{eqnarray}
\sigma_Q = \lim_{R\xrightarrow{}\infty}4\pi R^2 \frac{\langle\hat{I_s}(\textbf{r},t)\rangle}{\langle\hat{I_i}(\textbf{r},t)\rangle},
\end{eqnarray}
where $R$ is the distance between the target and the receiver. $\langle\hat{I_s}(\textbf{r},t)\rangle$ and $\langle\hat{I_i}(\textbf{r},t)\rangle$ denote the expectation values of the scattered and incident intensities at position $\textbf{r}$ and time $t$, respectively.
Using the Hamiltonian of atom-field interaction and the Weisskopf-Wigner theory \cite{Zubairy1997}, the expectation value of the scattered intensity is derived as
\begin{eqnarray}
\langle\hat{I}_s(\textbf{r},t)\rangle = \frac{\mathcal{E}_0^2}{2\eta N}|\sum^N_{n=1}(\textbf{d}_{ab}\cdot\textbf{e}_{\textbf{k}})e^{i\textbf{k}\cdot \Delta R_n}|^2,
\end{eqnarray}
where $\mathcal{E}_0 = \frac{e\omega^3}{8\pi^2c^3\epsilon_0}$, $N$ is the number of atoms, $\Delta R_n$ is the difference between the position vector of the $n$th atom and the receiver, $\eta$ represents the phase delay associated with the atomic transition, $\mathbf{d}_{ab}$ is the electric dipole moment of the atom consisting of ground state $a$ and excited state $b$, and $\mathbf{e}_{\mathbf{k}}$ is the polarization vector of the incident photon.
Assuming that the atoms have independent and identically distributed random dipole moment orientations, we obtain that the term $|(\textbf{d}_{ab}\cdot\textbf{e}_{\textbf{k}})|^2$ becomes a constant value and can be ignored. Using the law of energy conservation, we can obtain $\langle\hat{I_i}(\textbf{r},t)\rangle$ and then the single-photon QRCS is given by
\begin{eqnarray}
\sigma_Q = \frac{4 \pi A_{\perp}(\theta_i,\phi_i)|\sum_{n=1}^{N}e^{i(\textbf{k}_i-\textbf{k}_s)\cdot \textbf{x}_n}|^2}{\int_0^{2\pi}\int_0^{\pi/2}|\sum_{n=1}^{N}e^{i(\textbf{k}_i-\textbf{k}_s)\cdot \textbf{x}_n}|^2 \sin\theta_s d\theta_sd\phi_s},
\end{eqnarray}
where $A_{\perp}(\theta_i,\phi_i)$ is the projected cross-sectional area of the target at the incident angle $(\theta_i,\phi_i)$. $\textbf{k}_i$ and $\textbf{k}_s$ are the wave vectors of the incident and scattered photons, respectively. Furthermore, M-photon QRCS is also derived as :
\begin{eqnarray}
    \sigma_Q = \frac{4 \pi A_{\perp}(\theta_i,\phi_i)|\sum_{n=1}^{N}e^{i(\textbf{k}_i-\textbf{k}_s)\cdot \textbf{x}_n}|^{2M}}{\int_0^{2\pi}\int_0^{\pi/2}|\sum_{n=1}^{N}e^{i(\textbf{k}_i-\textbf{k}_s)\cdot \textbf{x}_n}|^{2M} \sin\theta_s d\theta_sd\phi_s},
\end{eqnarray}
where each photon operates separately. The detailed derivations are provided in the reference \cite{brandsema2017formulation}.

We utilize a two-photon entangled state as the incident field. To derive a modified QRCS formula, first, we describe the scattered intensity $\langle\hat{I_s}\rangle$ by analogy with the single-photon case. Then, we adopt the scattering theory of two-photon state \cite{Schotland:16}, where the incident field is assumed to be a maximally entangled state represented by 
$\mathcal{A}_i = \delta(\hat{\mathbf{k}}_1-\hat{\mathbf{k}}_2)$, with $\hat{\mathbf{k}}_1$ and $\hat{\mathbf{k}}_2$ being a unit wave vector of each incident photon. 
Based on the approach, we can write the two-photon amplitude with the T-matrix as :
\begin{eqnarray}\label{define A_s}
    \mathcal{A}_s(\hat{\mathbf{k}},\hat{\mathbf{k}}') = \int d\hat{\mathbf{k}}_1d\hat{\mathbf{k}}_2 \bra{\mathbf{k}}T\ket{\mathbf{k}_1}\bra{\mathbf{k}'}T\ket{\mathbf{k}_2}\mathcal{A}_i(\hat{\mathbf{k}}_1,\hat{\mathbf{k}}_2). \nonumber\\
\end{eqnarray}
Then we consider a collection of identical smaller scatters with a two-photon maximally entangled state. Inserting the momentum-space $T$-matrix elements of $\bra{\mathbf{k}}T\ket{\mathbf{k}'} = \sum_{a,b}t_{ab}(k)e^{i(\mathbf{k}\cdot\mathbf{r}_a-\mathbf{k}'\cdot\mathbf{r}_b)}$ and $\mathcal{A}_i = \delta(\hat{\mathbf{k}}_1-\hat{\mathbf{k}}_2)$ into Eq.~(\ref{define A_s}), thus, we derive the amplitude of scattered two-photon state as follows: 
\begin{eqnarray}
        \mathcal{A}_s(\hat{\mathbf{k}},\hat{\mathbf{k}}') =& 4\pi \sum_{a,b}\sum_{a',b'}t_{ab}(k) t_{a'b'}(k')e^{i(k\hat{\mathbf{k}}\cdot\mathbf{r}_a+k'\hat{\mathbf{k}}'\cdot\mathbf{r}_{a'})}\nonumber\\
                 &\times\operatorname{sinc}(|k\mathbf{r}_b+k'\mathbf{r}_{b'}|).
\end{eqnarray} 
By applying the results and following the same procedure as the single-photon QRCS derivation, the QRCS of a maximally entangled two-photon state is obtained as:
\begin{widetext}
    \begin{equation} \label{QRCS_two}
        \sigma_Q = \frac{4 \pi A_{\perp}(\theta_i,\phi_i)|\sum_{a,b,a',b'}t_{ab}(k)t_{a'b'}(k')e^{i(k\hat{\mathbf{k}}\cdot\mathbf{r}_a+k'\hat{\mathbf{k}}'\cdot\mathbf{r}_{a'})}\operatorname{sinc}(|k\mathbf{r}_b+k'\mathbf{r}_{b'}|)|^2}{\int_0^{2\pi}\int_0^{\pi/2}|\sum_{a,b,a',b'}t_{ab}(k)t_{a'b'}(k')e^{i(k\hat{\mathbf{k}}\cdot\mathbf{r}_a+k'\hat{\mathbf{k}}'\cdot\mathbf{r}_{a'})}\operatorname{sinc}(|k\mathbf{r}_b+k'\mathbf{r}_{b'}|)|^2 \sin\theta_s d\theta_sd\phi_s}.
    \end{equation}
\end{widetext}
Furthermore, we extend the analysis to two-photon entangled states with an arbitrary degree of entanglement using the double-Gaussian approximation. This parametrization enables control of the degree of entanglement through the pump divergence and phase-matching angular bandwidth. The incident state is represented as
\begin{eqnarray}
    \mathcal{A}_i(\hat{\mathbf{k}}_1,\hat{\mathbf{k}}_2)= N_s e^{-\frac{(\hat{\mathbf{k}}_{1\perp}+\hat{\mathbf{k}}_{2\perp})^2}{\sigma^2}}e^{-\frac{(\hat{\mathbf{k}}_{1\perp}-\hat{\mathbf{k}}_{2\perp})^2}{\mu^2}}, 
\end{eqnarray}
where $\frac{\sigma}{\sqrt{2}}$ denotes the pump divergence, $\frac{\mu}{\sqrt{2}}$ denotes the phase-matching angular bandwidth, and $N_s$ is a normalization constant. The subscript $\perp$ denotes transverse vector component. The corresponding Schmidt number, which quantifies the degree of entanglement, is given by $K = \frac{\sigma^2+\mu^2}{2\sigma\mu}$ \cite{PhysRevA.83.060302, Fedorov_2009}.
Here, we assume that the pump beam propagates along the $z$-axis, while the two-dimensional target lies in the $x$-$y$ plane. By substituting this incident state into Eq.~(\ref{define A_s}), the QRCS can be derived as :
\begin{widetext}
    \begin{align} \label{QRCS_arbit}
        \sigma_Q = \frac{4 \pi A_{\perp}(\theta_i,\phi_i)|\sum_{a,b,a',b'}t_{ab}(k)t_{a'b'}(k')e^{i(k\hat{\mathbf{k}}\cdot\mathbf{r}_a+k'\hat{\mathbf{k}}'\cdot\mathbf{r}_{a'})}I(k,k',\mathbf{r}_b,\mathbf{r}_b')|^2}{\int_0^{2\pi}\int_0^{\pi/2}\sum_{a,b,a',b'}t_{ab}(k)t_{a'b'}(k')e^{i(k\hat{\mathbf{k}}\cdot\mathbf{r}_a+k'\hat{\mathbf{k}}'\cdot\mathbf{r}_{a'})}I(k,k',\mathbf{r}_b,\mathbf{r}_b')|^2 \sin\theta_s d\theta_sd\phi_s},
    \end{align}
\end{widetext}
where
 \begin{eqnarray}
        I(k,k',\mathbf{r}_b,\mathbf{r}_b') &=& 4\pi^2\int d\theta_1d\theta_2 \sin\theta_1 \sin\theta_2 e^{-\alpha(\sin^2\theta_1+\sin^2\theta_2)} \nonumber \\
        &\times&\sum_{m=-\infty}^{\infty}(-1)^mI_m(-2\beta \sin \theta_1 \sin \theta_2) \nonumber\\
        &\times&J_m(k\rho_b\sin\theta_1)J_m(k'\rho_{b'}\sin\theta_2)e^{im(\varphi_b-\varphi_{b'})}, \label{arbitraryDOE}\nonumber
 \end{eqnarray}
and $\alpha = (\frac{1}{\sigma^2}+\frac{1}{\mu^2}),~ \beta = (\frac{1}{\sigma^2}-\frac{1}{\mu^2}), ~\mathbf{r}_b = (\rho_b\cos\varphi_b,\rho_b\sin\varphi_b), ~\mathbf{r}_{b'} = (\rho_{b'}\cos\varphi_{b'},\rho_{b'}\sin\varphi_{b'}) $.
    
Here, $J_m$ and $I_m$ denote the $m$-th order Bessel function and the $m$-th order modified Bessel function of first kind, respectively. Detailed derivation is provided in Appendix.

\section{Simulation results}
By using the Eq.~(\ref{QRCS_two}), at $\phi_i=0$, we conduct numerical simulations for a few two-dimensional targets with simple geometries and compare the results with those of single-photon QRCS and separable two-photon QRCS.
Specifically, we consider three types of two-dimensional target geometries, such as (i) square, (ii) circle, and (iii) triangle in monostatic and bistatic configurations.
Then, instead of the maximally two-photon entangled state,  we take an arbitrary degree of entanglement with the Eq. (\ref{QRCS_arbit}) for a square target in monostatic configuration, which exhibits a similar behavior to the other target geometries.

\subsection{Monostatic QRCS}
In the monostatic radar configuration, the transmitter and receiver are aligned in identical direction. Accordingly, the monostatic QRCS is defined as the radar cross section evaluated at the scattering angle equal to the incident angle $\theta$. 
The two-dimensional targets are assumed to consist of atoms arranged on a uniform grid. The square target is composed of a $31$ by $31$ atom array, while the circular and triangular targets are constructed by sampling atoms within the square grid and satisfy the respective geometric constraints. For each target, the number of atoms is set to $961$, $709$, and $462$, respectively.
The number of sampling points for the viewing angle $\theta$ is set to $400$. 

\begin{figure}[ht!]
    \centering
    \includegraphics[width=8cm]{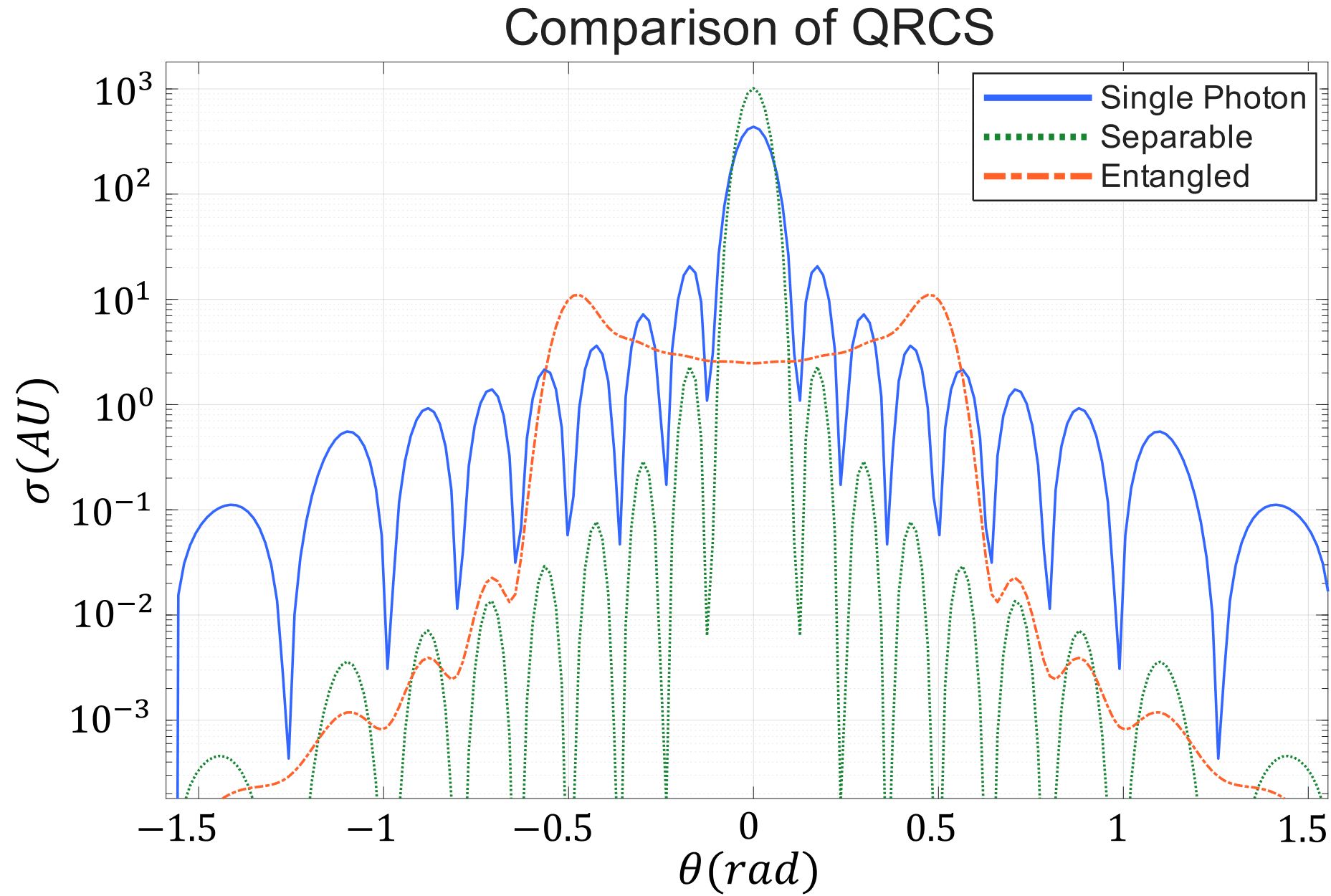}
    \caption{Monostatic QRCS for a square plate target consisting of 961 atoms arranged on a uniform grid.}
    \label{square_4lambda}
\end{figure}

\begin{figure}[ht!]
    \centering
    \includegraphics[width=8cm]{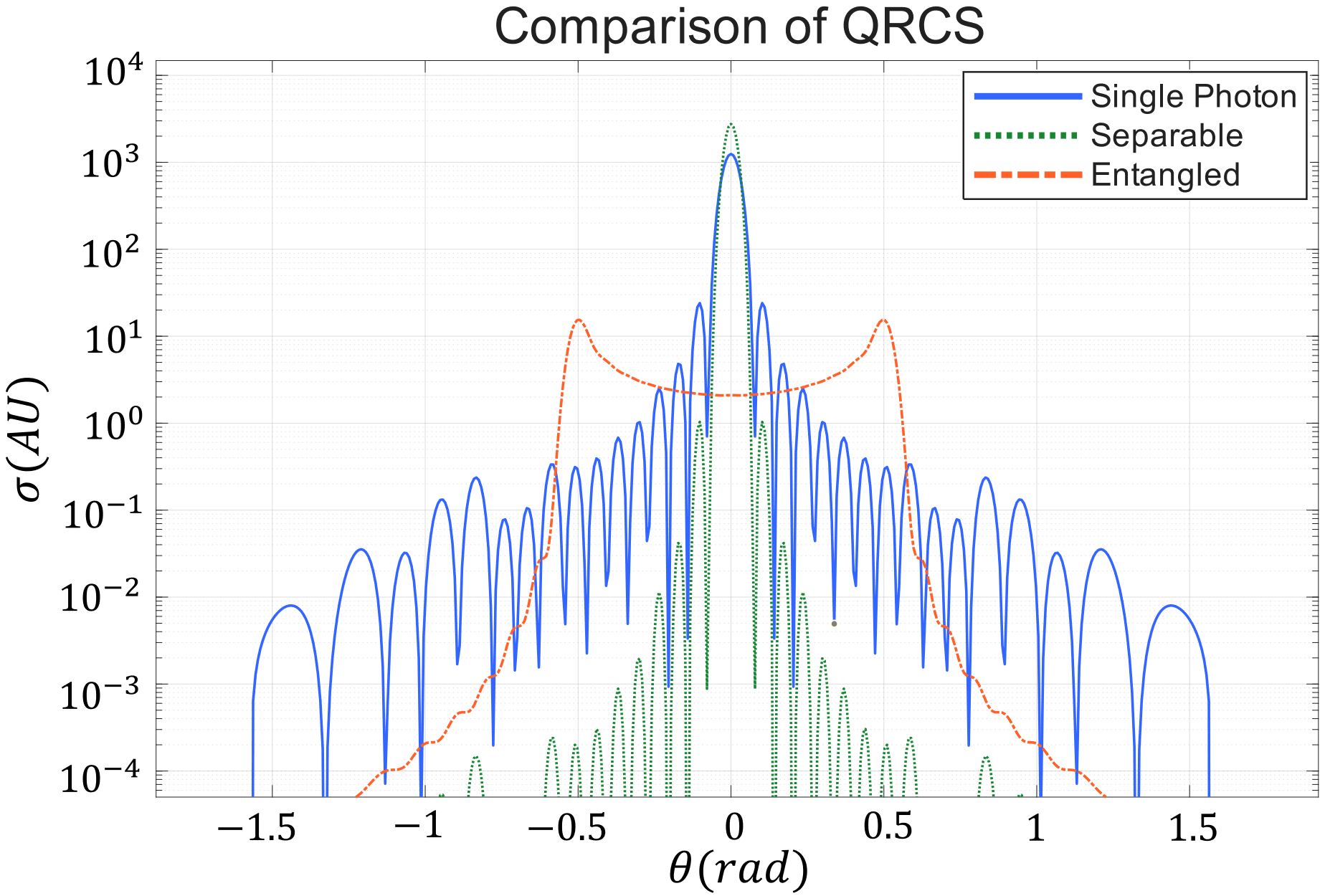}
    \caption{Monostatic QRCS for a circular plate target consisting of 709 atoms.}
    \label{circle_4lambda}
\end{figure}

\begin{figure}[ht!]
    \centering
    \includegraphics[width=8cm]{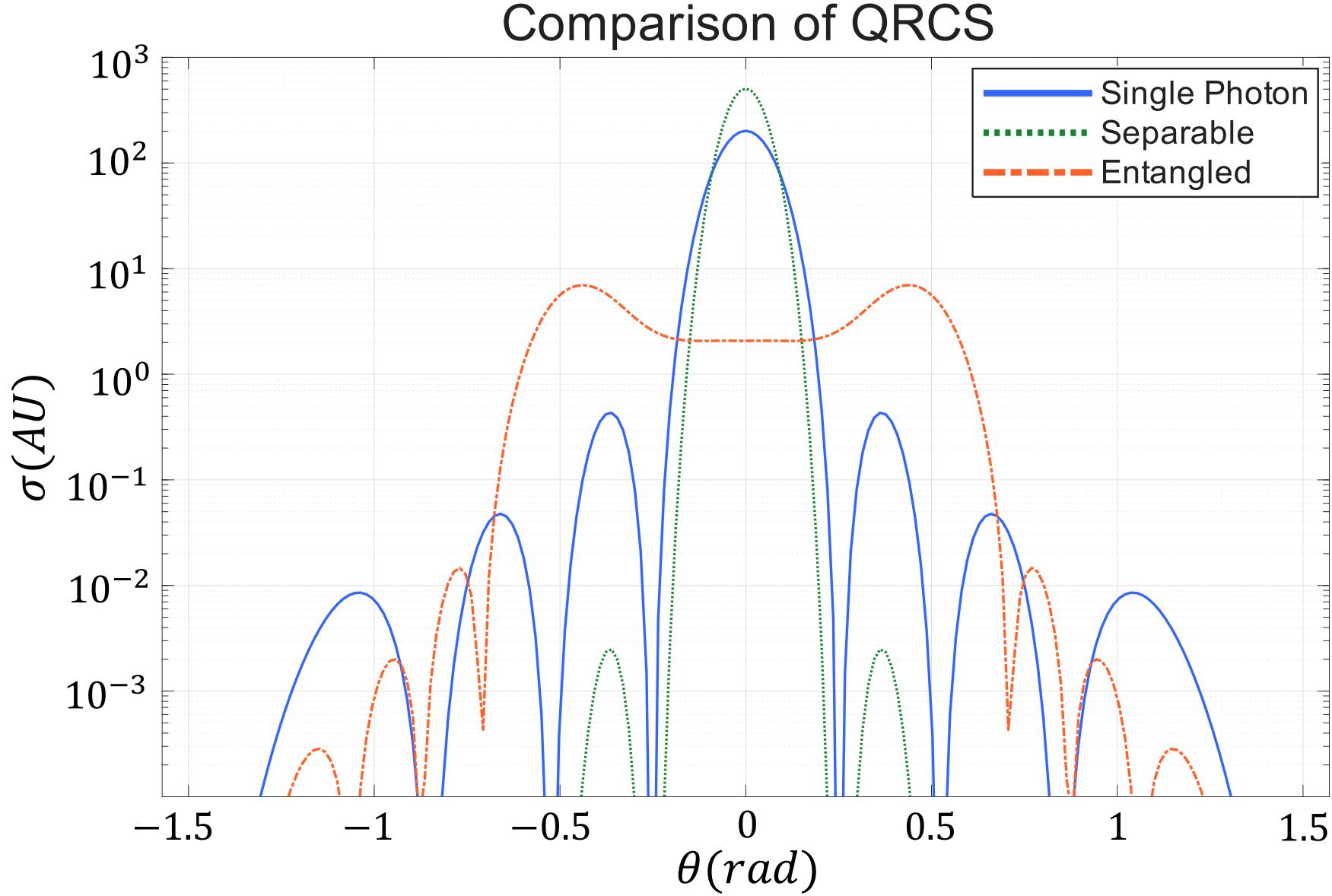}
    \caption{Monostatic QRCS for a triangular plate target consisting of 462 atoms.}
    \label{triangle_4lambda}
\end{figure}

Figures.~\ref{square_4lambda}--\ref{triangle_4lambda} show the results under the condition of $L=4\lambda$ and $L=1m$. 
We find that the scattering pattern of QRCS with two-photon entangled states differs from that obtained with single photon or separable two-photon states. 
In particular, QRCS with two-photon entangled states can offer an enhancement in a range of around $\pm 0.5$ radians rather than the main lobe that is equal to $0$ radian.
The main lobe shows a central dip. 
The side-lobe enhancement is the most pronounced in the circular and triangular targets which are smaller than the square target. 
Mathematically, this behavior is explained by the $\operatorname{sinc}$-dependent terms of the Eq.~(\ref{QRCS_two}).
The similar behaviors are also obtained in the different lengths of $L$ and the relations of $L=m\lambda$ ($m>1$). 
The enhancement is affected by the ratio between $L$ and $\lambda$.
When $m$ increases, the main lobe becomes narrower with more side-lobes.
Moreover, when we change the angle $\phi_i$, there is a slight change between $\pm 0.5$ and $\pm 1.5$ radians but it does not affect the overall trend of our results. 


\begin{figure}[ht!]
    \centering
    \includegraphics[width=8cm]{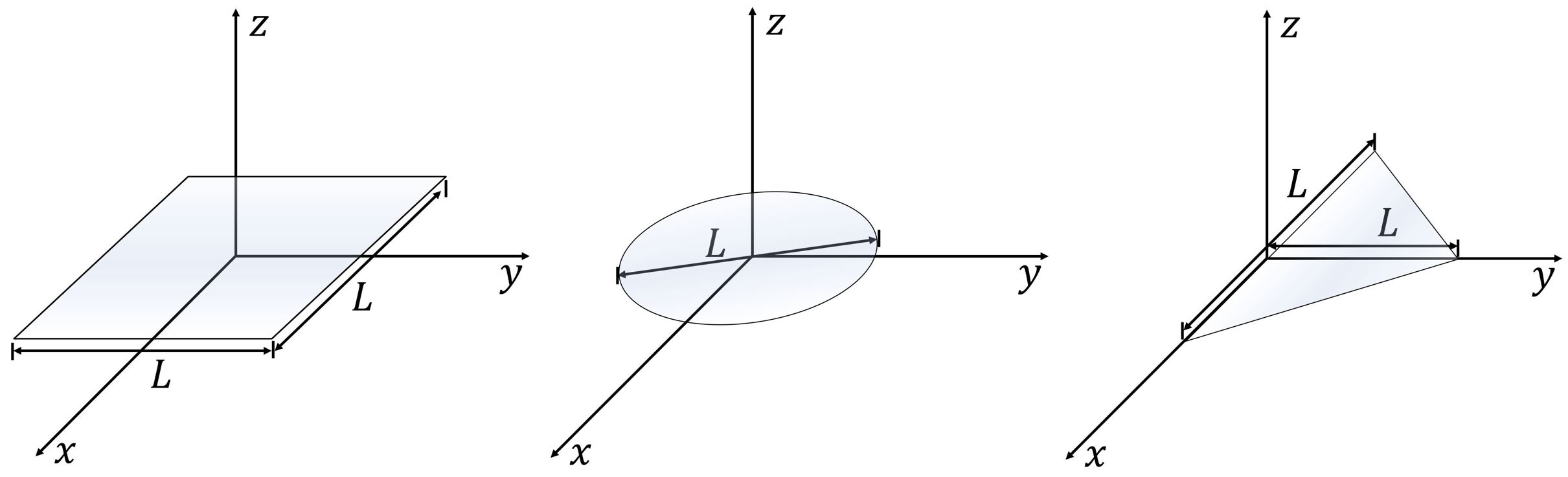}
    \caption{Targets with two-dimensional geometries. L denotes the length of a side of a square, the diameter of a circle, the base and height of a triangle. We set $L=1m$.}
    \label{2D_targets}
\end{figure}

\subsection{Bistatic QRCS}

\begin{figure}[ht!]
    \centering
    \includegraphics[width=8cm]{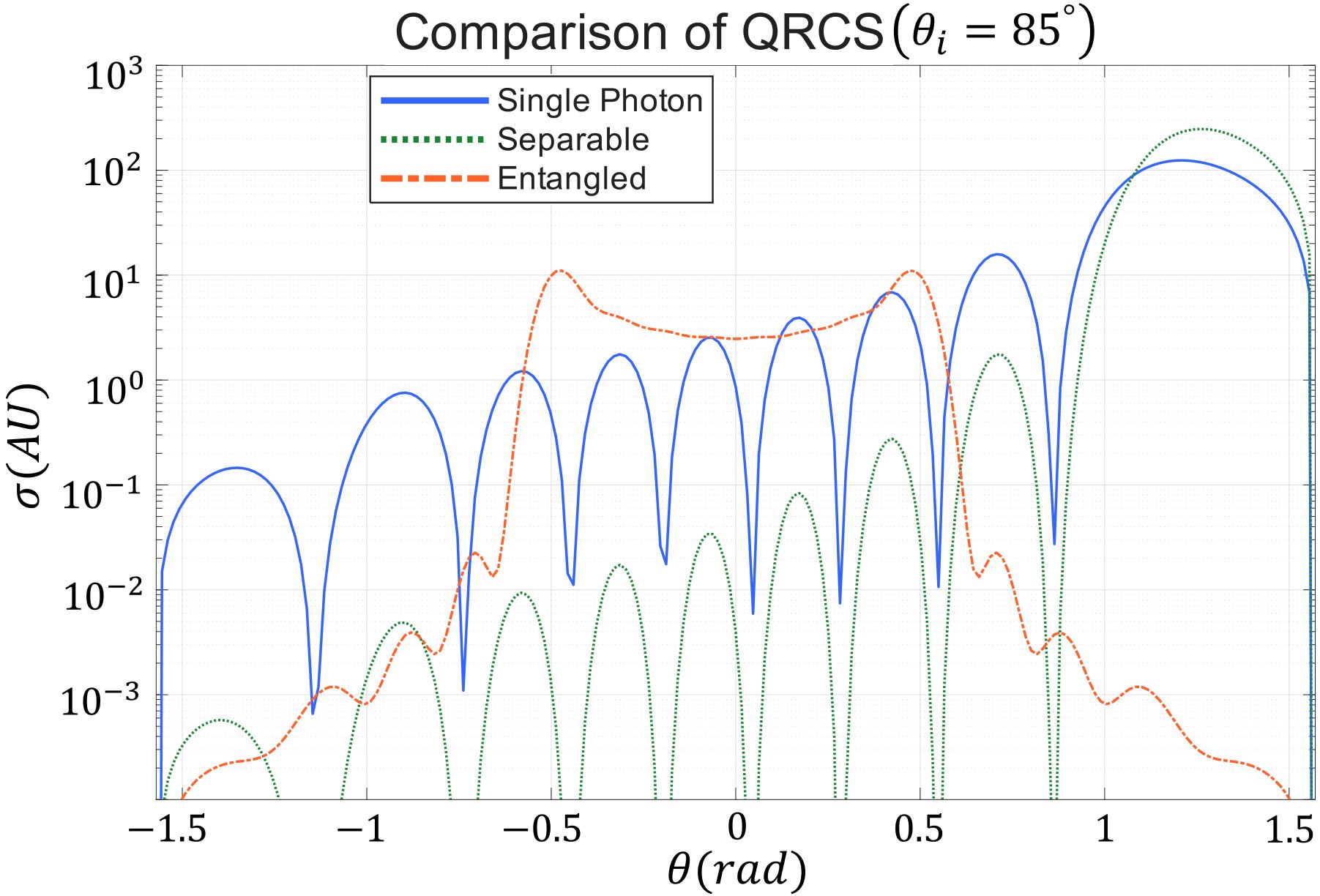}
    \caption{Bistatic QRCS for a square plate target at an incident angle $\theta_i=85^{\circ}$.}
    \label{square_bi_85}
\end{figure}

\begin{figure}[ht!]
    \centering
    \includegraphics[width=8cm]{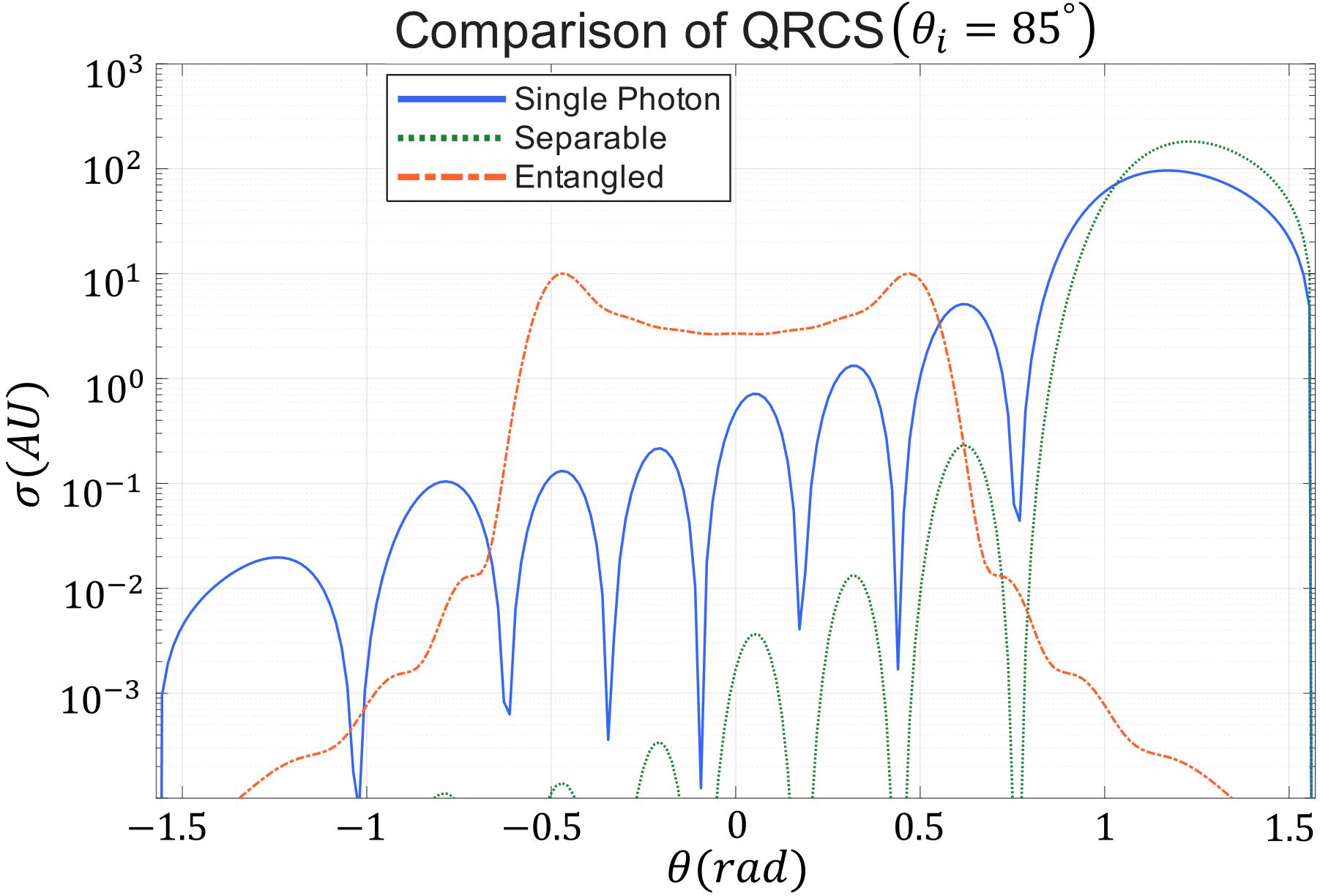}
    \caption{Bistatic QRCS for a circular plate target at an incident angle $\theta_i=85^{\circ}$.}
    \label{circle_bi_85}
\end{figure}

\begin{figure}[ht!]
    \centering
    \includegraphics[width=8cm]{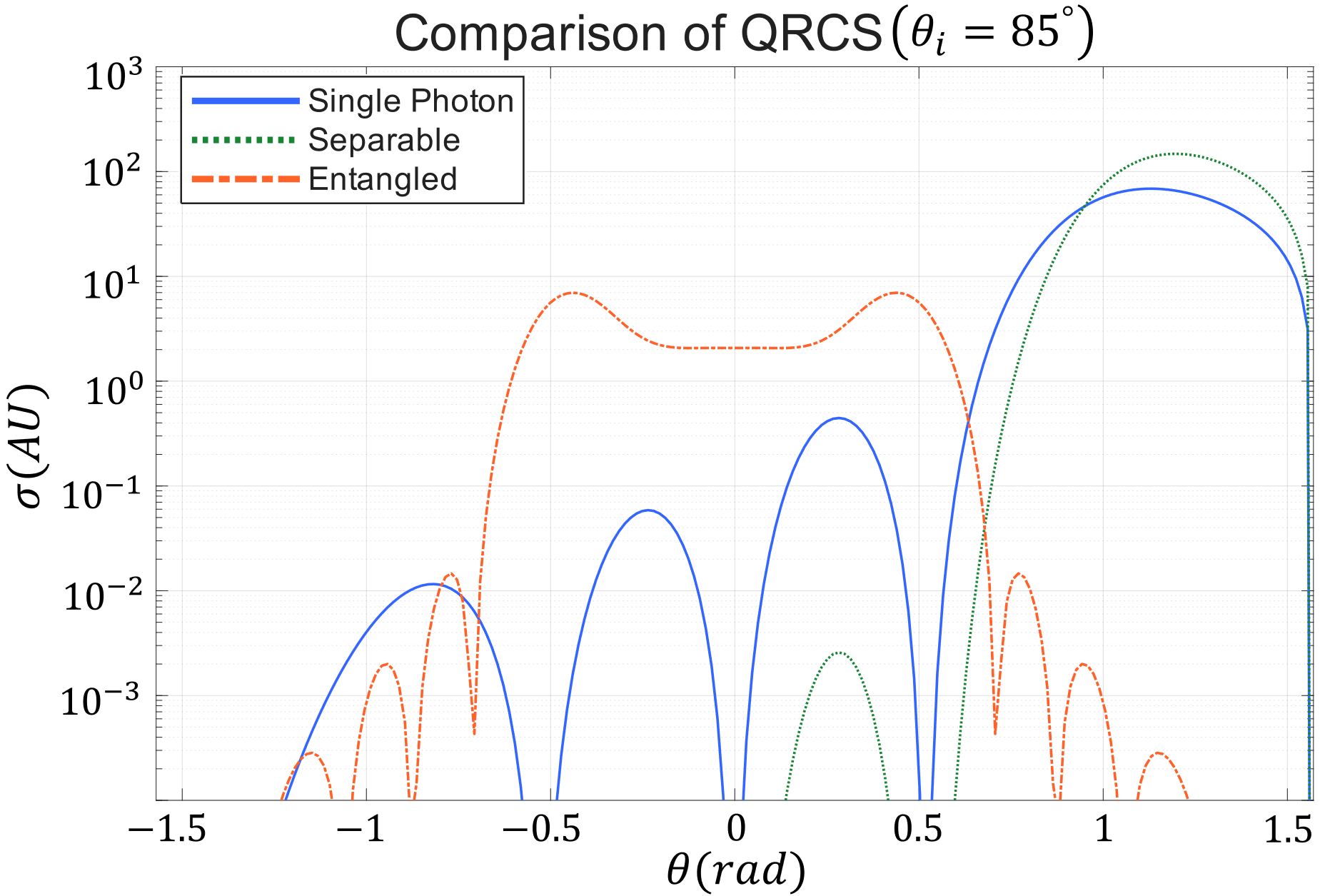}
    \caption{Bistatic QRCS for a triangular plate target at an incident angle $\theta_i=85^{\circ}$.}
    \label{triangle_bi_85}
\end{figure}

The bistatic QRCS considers the scenarios where the transmitter and receiver are spatially separated, so the scattered field is measured at a distinct angle from the incident angle. Accordingly, the bistatic QRCS is defined as the radar cross section evaluated over all scattering directions under a fixed incident angle $\theta_i$.
In this subsection, we compare bistatic QRCS for different target geometries with $L = 4\lambda$ and $L=1m$. 
Since our formulation utilizes a maximally entangled incident state represented by $\mathcal{A}_i = \delta(\hat{\mathbf{k}}_1-\hat{\mathbf{k}}_2)$, 
the corresponding QRCS reflects a scattering response over all scattering directions that integrates over the angular uncertainty of the entangled incident state. 
Thus, the QRCS is independent of the incident angle and the scattering pattern of the maximally entangled state remains unchanged. 
As the incident angle varies, the lobes of single-photon and separable two-photon states shifted accordingly. 
In Figs.~\ref{square_bi_85}--\ref{triangle_bi_85}, we compare the results for bistatic QRCS with incident angle $\theta_i = 85^{\circ}$. 
The similar behaviors are also obtained in the different lengths of $L$ and the relations of $L=m\lambda$ ($m>1$).


\subsection{QRCS with an arbitrary degree of entanglement}
\begin{figure}[ht!]
    \centering
    \includegraphics[width=8cm]{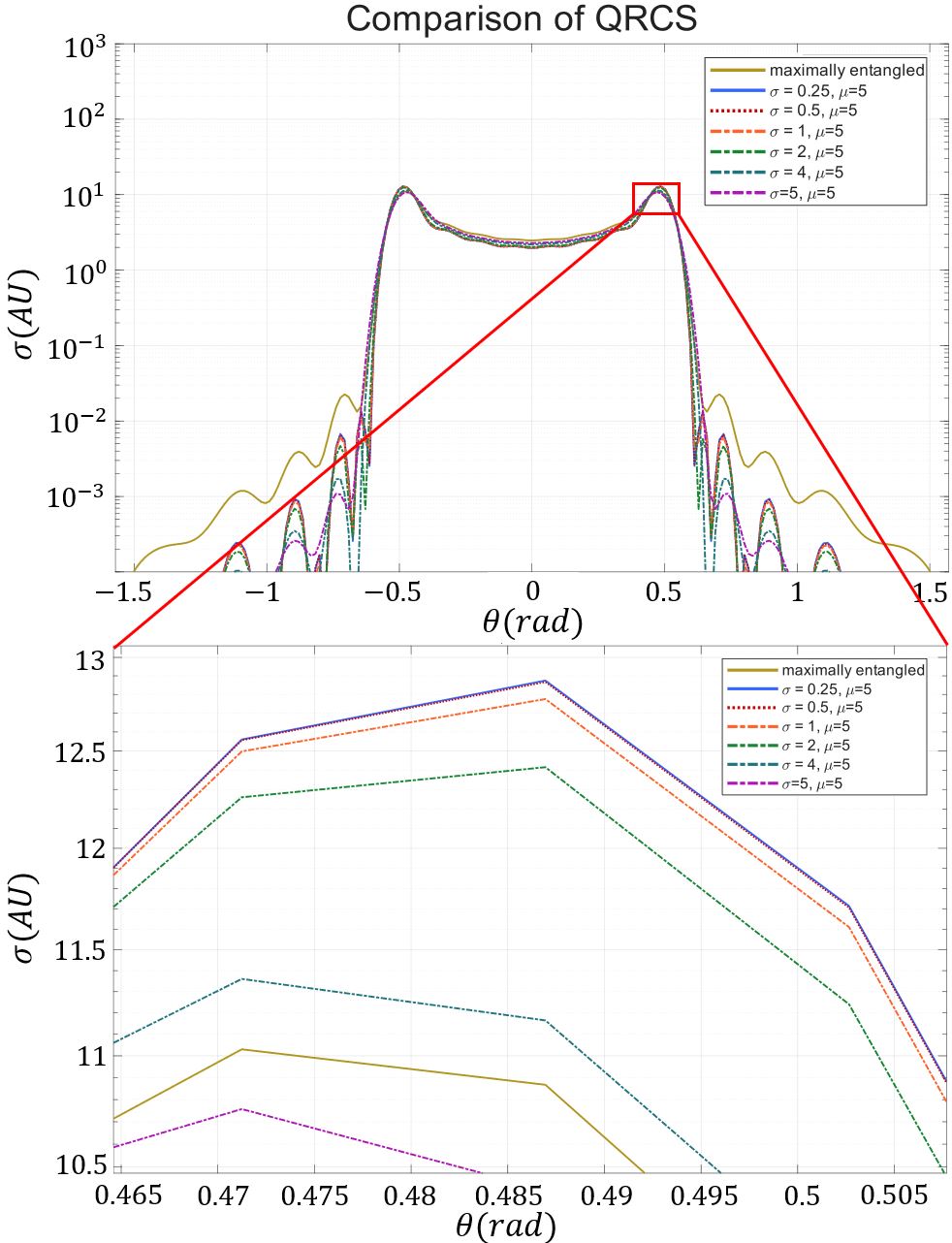}
    \caption{QRCS with an arbitrary degree of entanglement for a square plate target as a function of scattering angle $\theta$,
    where $\mu=5$ is fixed and $\sigma=0.25,~0.5,~1,~2,~4,~5$.}
    \label{DOE_total}
\end{figure}

In the previous section, we derived the general form of the QRCS for a two-photon entangled state with an arbitrary degree of entanglement using the double-Gaussian approximation. Using the the Eq. (\ref{QRCS_arbit}), we conduct numerical simulations and compare the scattering patterns for a square plate target with $L = 4\lambda$ and $L=1m$. 
In Fig.~\ref{DOE_total}, we compare the QRCS with different values of $\sigma$ at $\mu=5$.
We obtain that non-maximal (or partially) entangled states can ourperform the maximal entangled state. 
In the inset of Fig.~\ref{DOE_total}, on average, the non-maximal entangled state (at $\sigma=0.25,~\nu=5$) can achieve up to a $14.67\%$ enhancement over the maximally entangled state.
Unlike the case of the maximal entangled state that admits a closed-form expression, the QRCS with states having an arbitrary degree of entanglement involves an infinite series. To perform the numerical simulation, we approximate the QRCS by truncating the infinite sum at the $20$th term, owing to computational limitations. 


At $\sigma=5$ and $\mu=5$, the two-photon state corresponds to a separable two-photon state. However, its scattering pattern differs from that of the conventional separable two-photon state represented by the green curve in Fig.~\ref{square_4lambda}. This discrepancy arises because the conventional separable two-photon state assumes a distinguishable two-photon, whereas our formulation considers an indistinguishable two-photon state. 
As the degree of entanglement increases, naturally, the scattering patterns approach those of the maximally entangled state. 

\section{Conclusion}
We derived a modified QRCS formula with the maximally entangled two-photon state. 
We performed numerical simulations to compare the resulting QRCS with those obtained from single-photon and separable two-photon states for various two-dimensional target geometries, including square, circular, and triangular shapes in both monostatic and bistatic configurations.
 We obtained that QRCS with two-photon entangled states can provide an enhancement in side lobes rather than the main-lobe regime. 
Specifically, at a scattering angle of about $\pm 0.5$ radians, the two-photon entangled state pronouncedly outperforms the counterpart states.
Thus, it can be a novel strategy to detect the target at the scattering angle, $\pm 0.5$ radians.
 We note that the classical RCS is not considered a benchmark since single-photon QRCS took an advantage in the side-lobe compared to the classical RCS \cite{Brandsema}.
Thus, it is natural to say that the QRCS with two-photon entangled state can offer side-lobe enhancement compared to the classical RCS.
 In addition, since the assumed incident state does not specify a definite incident direction, the resulting QRCS is robust in bistatic configurations, whereas the scattering patterns of conventional QRCS shift with the incident angle.
 
Employing the double-Gaussian approximation, then, we extended the formulation to describe QRCS with a two-photon entangled state having an arbitrary degree of entanglement.
By conducting numerical simulations truncated at the $20$th term, for a square plate target, we obtained that non-maximal two-photon entangled states can ourperform the maximal two-photon entangled state.  
Although the numerical simulation was conducted with finite terms, the scattering pattern converges to the maximal two-photon entangled state in the limit of an increasing entanglement. 


\begin{acknowledgments}
This work was supported by Defense Acquisition Program Administration, the Agency for Defense Development, and the NIWC Pacific Naval Innovative Science \& Engineering (NISE) program.

\end{acknowledgments}

\newpage
\appendix
\onecolumngrid
\section{Derivation of QRCS using entangled two-photon state with arbitrary degree of entanglement}
In this section, we provide detailed derivation of Eq.~(\ref{arbitraryDOE}). Our incident state is assumed to be approximated as double-Gaussian form, as follows :
\begin{equation}
    \mathcal{A}_i(\hat{\mathbf{k}}_1,\hat{\mathbf{k}}_2)= N_s e^{-\frac{(\hat{\mathbf{k}}_{1\perp}+\hat{\mathbf{k}}_{2\perp})^2}{\sigma^2}}e^{-\frac{(\hat{\mathbf{k}}_{1\perp}-\hat{\mathbf{k}}_{2\perp})^2}{\mu^2}}.
\end{equation}
According to the scattering theory of entangled two-photon state, the scattered amplitude of two-photon state about multiple scatterer is given as:
\begin{equation}
    \mathcal{A}_s(\hat{\mathbf{k}},\hat{\mathbf{k}}') = \int d\hat{\mathbf{k}}_1d\hat{\mathbf{k}}_2 \bra{\mathbf{k}}T\ket{\mathbf{k}_1}\bra{\mathbf{k}'}T\ket{\mathbf{k}_2}\mathcal{A}_i(\hat{\mathbf{k}}_1,\hat{\mathbf{k}}_2),
\end{equation}
where $\bra{\mathbf{k}}T\ket{\mathbf{k}_1} = \sum_{a,b}t_{ab}(k)e^{i(\mathbf{k}\cdot\mathbf{r}_a-\mathbf{k}_1\cdot\mathbf{r}_b)}$ and $\bra{\mathbf{k'}}T\ket{\mathbf{k}_2} = \sum_{a',b'}t_{a'b'}(k')e^{i(\mathbf{k'}\cdot\mathbf{r}_{a'}-\mathbf{k_2}\cdot\mathbf{r}_{b'})}$. By combining these equations, 
\begin{equation}
    \mathcal{A}_s(\hat{\mathbf{k}},\hat{\mathbf{k}}') = N_s\sum_{a,b,a',b'}t_{ab}(k)t_{a'b'}(k')e^{i(k\hat{\mathbf{k}}\cdot\mathbf{r}_a+k'\hat{\mathbf{k}}'\cdot\mathbf{r}_{a'})}I(k,k',\mathbf{r}_b,\mathbf{r}_b'),
\end{equation}
where
\begin{equation}
    I(k,k',\mathbf{r}_b,\mathbf{r}_b') = \int d\hat{\mathbf{k}}_{1}d\hat{\mathbf{k}}_{2} e^{-\frac{(\hat{\mathbf{k}}_{1\perp}+\hat{\mathbf{k}}_{2\perp})^2}{\sigma^2}}e^{-\frac{(\hat{\mathbf{k}}_{1\perp}-\hat{\mathbf{k}}_{2\perp})^2}{\mu^2}}e^{-i(k\hat{\mathbf{k}}_1\cdot\mathbf{r}_b+k'\hat{\mathbf{k}}_2\cdot\mathbf{r}_{b'})}.
\end{equation} 
Since we assume that the pump beam propagates along $z$-axis, we can write $\hat{\mathbf{k}}_{j\perp} = (\sin\theta_j\cos\phi_j, \sin\theta_j \sin\phi_j).$ Then we can write
\begin{equation}
    e^{-\frac{(\hat{\mathbf{k}}_{1\perp}+\hat{\mathbf{k}}_{2\perp})^2}{\sigma^2}}e^{-\frac{(\hat{\mathbf{k}}_{1\perp}-\hat{\mathbf{k}}_{2\perp})^2}{\mu^2}} = e^{-\alpha(\sin^2\theta_1+\sin^2\theta_2-2\beta\sin\theta_1\sin\theta_2\cos(\phi_1-\phi_2))},
\end{equation}
where $\alpha = (\frac{1}{\sigma^2}+\frac{1}{\mu^2}),\beta = (\frac{1}{\sigma^2}-\frac{1}{\mu^2}).$
Let the position of each scatterer $\mathbf{r}_b = (\rho_b\cos\varphi_b,\rho_b\sin\varphi_b,0).$ Then 
\begin{equation}
    e^{-i(k\hat{\mathbf{k}}_1\cdot\mathbf{r}_b+k'\hat{\mathbf{k}}_2\cdot\mathbf{r}_{b'})} = e^{-i(k\rho_b\sin\theta_1\cos(\phi_1-\varphi_b)+k'\rho_{b'}\sin\theta_2\cos(\phi_2-\varphi_{b'}))}.
\end{equation}
Here, we adopt the Bessel function expansion as:
\begin{equation}
    e^{-2\beta\sin\theta_1\sin\theta_2\cos(\phi_1-\phi_2)} = \sum^{\infty}_{m=-\infty}I_m(-2\beta\sin\theta_1\sin\theta_2)e^{im(\phi_1-\phi_2)},
\end{equation}
and
\begin{equation}
    e^{-iA\cos(\phi-\varphi)} = \sum^{\infty}_{n=-\infty}J_n(A)e^{in(\phi-\varphi)}.
\end{equation}
Using its orthogonality, finally the integral is summarized as :
\begin{align}
    I(k,k',\mathbf{r}_b,\mathbf{r}_b') = 4\pi^2\int d\theta_1d\theta_2 &\sin\theta_1 \sin\theta_2 e^{-\alpha(\sin^2\theta_1+\sin^2\theta_2)} \nonumber \\
    &\sum_{m=-\infty}^{\infty}(-1)^mI_m(-2\beta \sin \theta_1 \sin \theta_2)J_m(k\rho_b\sin\theta_1)J_m(k'\rho_{b'}\sin\theta_2)e^{im(\varphi_b-\varphi_{b'})}
\end{align}

\end{document}